\documentclass{jfm}

\usepackage{graphicx}
\usepackage{epstopdf}
\usepackage{natbib}
\usepackage{amsmath}
\usepackage{amssymb}
\usepackage{MnSymbol}
\usepackage{float}

\usepackage[svgnames,x11names]{xcolor}
\usepackage[normalem]{ulem}

\ifCUPmtlplainloaded \else
  \checkfont{eurm10}
  \iffontfound
    \IfFileExists{upmath.sty}
      {\typeout{^^JFound AMS Euler Roman fonts on the system,
                   using the 'upmath' package.^^J}%
       \usepackage{upmath}}
      {\typeout{^^JFound AMS Euler Roman fonts on the system, but you
                   dont seem to have the}%
       \typeout{'upmath' package installed. JFM.cls can take advantage
                 of these fonts,^^Jif you use 'upmath' package.^^J}%
      }
  \else
  \fi
\fi

\ifCUPmtlplainloaded \else
  \checkfont{msam10}
  \iffontfound
    \IfFileExists{amssymb.sty}
      {\typeout{^^JFound AMS Symbol fonts on the system, using the
                'amssymb' package.^^J}%
       \usepackage{amssymb}%
         \let\leq=\leqslant
         
      }{}
  \fi
\fi

\ifCUPmtlplainloaded \else
  \IfFileExists{amsbsy.sty}
    {\typeout{^^JFound the 'amsbsy' package on the system, using it.^^J}%
     \usepackage{amsbsy}}
    {}
\fi

\newsavebox{\astrutbox}
\sbox{\astrutbox}{\rule[-5pt]{0pt}{20pt}}

\title[Transition to turbulence in pulsating pipe flow]{Transition to turbulence in pulsating pipe flow}

\author[Duo Xu, Sascha Warnecke, Baofang Song, Xingyu Ma and Bj{\"o}rn Hof]%
{Duo Xu$^{1,2,3}$ \thanks{Email address for correspondence: duo.xu@fau.de}, Sascha Warnecke$^{1,2}$, Baofang Song$^{2}$,  Xingyu Ma$^{2}$ and Bj\"{o}rn Hof$^{2}$}

\affiliation{$^1$ Max Planck Institute for Dynamics and Self-Organization, G{\"o}ttingen 37073, Germany\\[\affilskip]
$^2$ Institute of Science and Technology Austria, Am Campus 1, Klosterneuburg 3400, Austria\\[\affilskip]
$^3$ Friedrich-Alexander-Universit\"{a}t Erlangen-N\"{u}rnberg, Erlangen 91058, Germany}

\pubyear{?}
\volume{?}
\pagerange{?}
\date{?; revised ?; accepted ?. - To be entered by editorial office}
\begin{document}

\maketitle

\begin{abstract}
Fluid flows in nature and applications are frequently subject to periodic velocity modulations. Surprisingly, even for the generic case of flow through a straight pipe, there is little consensus regarding the influence of pulsation on the transition threshold to turbulence: while most studies predict a monotonically increasing threshold with pulsation frequency (i.e. Womersley number, $\alpha$), others observe a decreasing threshold for identical parameters and only observe an increasing threshold at low $\alpha$.  In the present study we apply recent advances in the understanding of transition in steady shear flows to pulsating pipe flow. For moderate pulsation amplitudes we find that the first instability encountered is subcritical  (i.e. requiring finite amplitude disturbances) and gives rise to localized patches of turbulence (``puffs'') analogous to steady pipe flow. By monitoring the impact of pulsation on the lifetime of turbulence we map the onset of turbulence in parameter space.
Transition in pulsatile flow can be separated into three regimes. At small Womersley numbers the dynamics are dominated by the decay turbulence suffers during the slower part of the cycle and hence transition is delayed significantly. As shown in this regime thresholds closely agree with estimates based on a quasi steady flow assumption only taking puff decay rates into account. The transition point predicted in the zero $\alpha$ limit equals to the critical point for steady pipe flow offset by the oscillation Reynolds number (i.e. the dimensionless oscillation amplitude). In the high frequency limit on the other hand puff lifetimes are identical to those in steady pipe flow and hence the transition threshold appears to be unaffected by flow pulsation. In the intermediate frequency regime the transition threshold sharply drops (with increasing $\alpha$) from the decay dominated (quasi steady) threshold to the steady pipe flow level.
\end{abstract}

\begin{keywords}
\end{keywords}

\section{Introduction}\label{sec:introduction}
Pulsatile flows are often at the verge of being turbulent. The most prominent example is cardiovascular flow, where Reynolds numbers in the larger blood vessels are in the transitional regime. Likewise hydrodynamic instabilities and turbulent wall-shear-stress fluctuations have been associated with certain cardiovascular diseases \citep{Freis64, Flaherty72, Stein76, Silver78, Bogren94}. While in physiological flows non-Newtonian fluid properties and the often complex geometry cause additional complications,  even the much simpler case of transition in pulsatile flow of a Newtonian fluid in a straight pipe is not well understood. In contrast to steady pipe flow which is governed solely by the Reynolds number and where recenlty much progress has been made towards understading the transition scenario\citep{Eckhardt07, Avila11, Barkley15}, for pulsating flows, in addition to the Reynolds number ($Re_s=U_sD/\nu$, where $U_s$ corresponds to the steady component of velocity $D$ is the pipe diameter and $\nu$ is the kinematic viscosity of the fluid), the amplitude ($A=U_o/U_s$, where $U_o$ is the oscillation component of velocity) and the pulsation frequency (usually expressed by the Womersley number $\alpha=D\sqrt{2\pi f/\nu}/2$) have to be taken into account.
Regarding the stability of pulsating flows in straight pipes, earlier studies reported flows to be linearly stable \citep{Kerczek74}, however a more recent study reported linear instability for large pulsation amplitudes \citep{Thomas11}. Nevertheless typically turbulence can be observed for values well below the predicted linear stability border and the transition points determined in the present study are exclusively in the linearly stable regime.

\begin{figure}
\centering
\includegraphics[width = 0.6\textwidth, trim=0cm 0cm 0cm 0cm, clip]{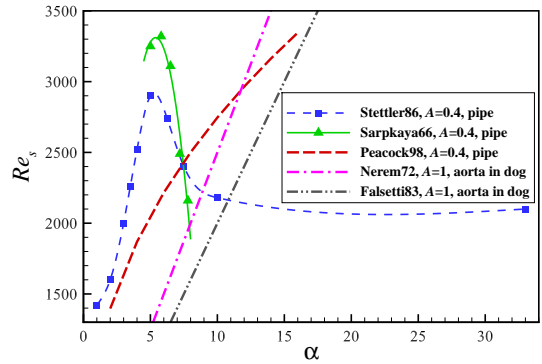}\\
\caption{\label{fig:former_exp} (Color online) Transition thresholds to turbulence reported in earlier studies of pulsatile pipe flow.}
\end{figure}

Transition thresholds for the onset of turbulence reported in earlier studies (shown in FIG. \ref{fig:former_exp}) do not only differ quantitatively, but also find opposite trends with Womersley number.
The majority of above studies \citep{Nerem72a,Nerem72b,Falsetti83,Peacock98} suggest that the threshold increases with Womersley number. While the first three studies were carried out in dog aortas, the latter result was obtained in a straight pipe ($200D$ in length) using a Newtonian fluid. \citet{Stettler86} on the other hand report an increasing trend (from a $330D$ pipe setup) only for low Womersley numbers whereas for $\alpha>5$  the transition point is found to decrease and eventually settles to a constant value for $\alpha >10$. A decreasing threshold for $5<\alpha <10$ had also been reported by \citet{Sarpkaya66} in $\sim 1000D$ pipes, however transition points deviate considerably from those of \citet{Stettler86}. The large Womersley number regime has also been investigated in a $150D$ pipe recently by \citet{Trip12} and although thresholds were not determined explicitly, transition (based on intermittency factor measurements) appeared to be unaffected by pulsation for $\alpha>10$. Regarding the effect on pulsation amplitude, \citet{Gilbrech63, Hershey68} found that the transition Reynolds number decreased when the pulsation amplitude (or {pulsation Reynolds number} $Re_o=U_oD/\nu$) was increased.

The discrepancies between above studies regarding the scaling with $\alpha$ may partially be related to the varying degrees of complexity of the geometries considered. Some studies investigated actual blood flows in the aorta, others were concerned with the simpler and more generic case of a Newtonian fluid in a straight rigid pipe. Nevertheless the increasing trend with Womersley number reported for aortic flows has been confirmed for pulsating pipe flow by \citet{Peacock98}. Another possible reason for the disagreement may lie in the initial preparation of flows: In some cases external perturbations were added to trigger turbulence \citep{Trip12, Stettler86, Sarpkaya66}, while others \citet{Nerem72a,Nerem72b,Falsetti83,Peacock98} simply increased parameters until turbulence first appeared without additional external triggers. The latter may appear closer to real blood flows, however from a hydrodynamic stability point of view these conditions are not well suited for studies of subcritical transition. The transition point in those cases intrinsically depends on the level of finite amplitude perturbations specific to the experimental set up (i.e. imperfections of the set up such as small steps at junctions between pipe sections, secondary flows created by the inlet etc.). Since these perturbation levels differ from experiment to experiment this ``natural transition point'' is not generic. 
 
The aim of the present study is to clarify the effect of pulsation on transition for the generic case of Newtonian fluid flow through a straight rigid pipe and to provide a measure of the transition threshold that is set up independent. For this purpose we will use lifetimes of turbulent puffs to quantify the effect of flow pulsation on the onset of sustained turbulence. In steady pipe flow at low Reynolds numbers puffs are the characteristic turbulent structures that appear if a sufficient perturbation is applied to the flow. Likewise for pulsating pipe flow, over the entire parameter regime investigated in the present study, puffs are the first turbulent structures to appear. Puffs have a typical length $\sim 20D$ and are advected downstream at approximately the mean flow speed. Recent studies of steady pipe flow showed that puffs are intrinsically transient and decay after long times following a memoryless process \citep{Hof06, Hof08, Avila10}. Turbulence eventually becomes sustained, i.e. it persists for indefinitely long pipes, due to a spatial spreading process that outweighs the decay rate of individual puffs \citep{Avila11}. In principle it would be possible to determine the onset of sustained turbulence in the same manner for pulsating flows. However this would require studies of the variation of decay and spreading processes with Reynolds number, Womersley number and pulsation amplitude, and the very long time scales of the underlying processes make this impractical. A close estimate of the transition threshold can be obtained from lifetimes alone and we assume that the critical point is closely approached once lifetimes of puffs are large. It should be noted that median lifetimes of puffs and puff survival probabilities do not depend on the specifics of the experimental set up and in the case of steady pipe flow lifetimes are a function of the Reynolds number only. Median puff lifetimes (for steady flow) have been determined in various different experimental set ups \citep{Hof08,Kuik10,Avila11,Samanta13} as well as in direct numerical simulations and were found to be in excellent quantitative agreement. In particular the characteristic lifetime of puffs (except for the initial formation time) are also independent of the type of perturbation used to trigger \citep{delozar09}. In pulsating flow for a meaningful measure of the transition threshold puffs during their passage through the pipe have to experience at least one pulsation cycle. Otherwise the measured lifetimes are phase dependent. Meaning that if a puff only experiences part of a cycle it experiences a mean Reynolds number different from that of the full cycle. In particular at low Womersley numbers insufficient pipe lengths (observation times) can lead to significant underestimates of the transition threshold.

\section{Experiments}\label{sec:experiment}
\begin{figure}
\centering
\includegraphics[width = 1\textwidth, trim=0cm 0cm 0cm 0cm, clip]{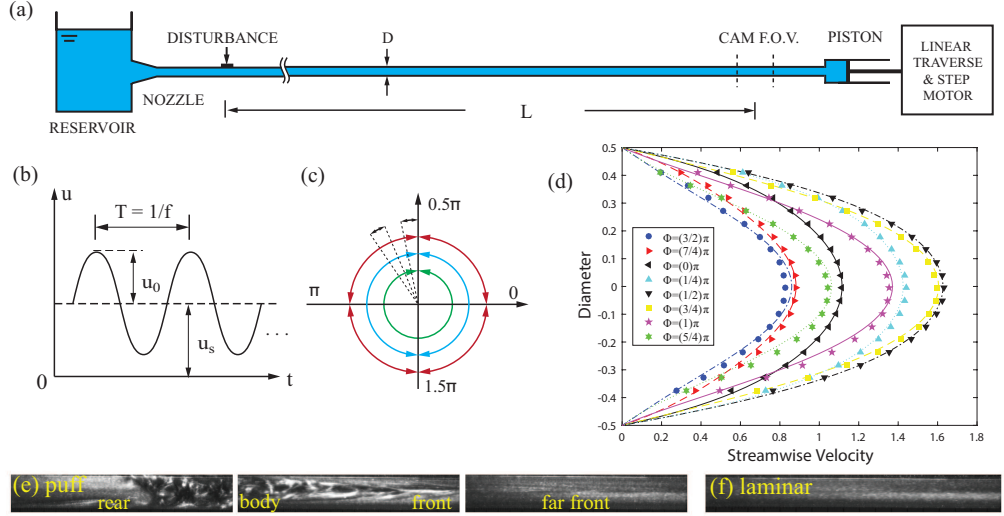}
\caption{\label{fig:facility} {(Color online) (a) Sketch of the pulsatile pipe flow setup (pipe-1: $D=10~\textrm{mm}$ and $L=350D$, pipe-2: $D=7.18~\textrm{mm}$ and $L=1300D$, pipe-3: $D=4~\textrm{mm}$ and $L=2250D$), (b) the sinusoidally modulated speed of the piston and (c) the phase covered by the injection perturbation. (d) Comparison of measured velocity profiles (symbols) and analytical solutions (lines in corresponding color) at $Re_s=2000$, $\alpha=5$ and $A=0.4$. Typical flow visualization images are shown for turbulent (e) and laminar (f) flow. (The images display a pipe segment of about $5D$ long and flow direction is from left to right.) }}
\end{figure}

\begin{figure}
\centering
\includegraphics[width = 1\textwidth, trim=0cm 0cm 0cm 0cm, clip]{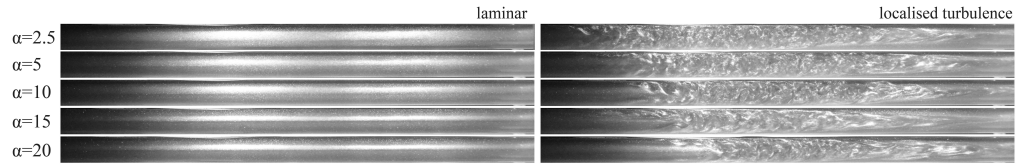}
\caption{\label{fig:snapshot} Visualization samples of laminar flow which is observed in the absence of external perturbations (left column). When perturbed upstream, turbulence could be excited locally (right column). The examples shown are for parameter values of $(\alpha; Re_s)$ of $(2.5; 3100), (5; 2800), (10; 2500), (15;2500), (20; 2500)$ from top to bottom. The flow direction is from left to right.}
\end{figure}

Experiments were carried out in straight, rigid pipes of circular cross section. Three set ups were used which mainly differed in the pipe diameter and the pipe length. In the first case the pipe is composed of five glass tubes resulting in a total length of $5.5~\textrm{m}$. Tubes have an inner diameter of $D=10 \pm 0.01~\textrm{mm}$ and hence a dimensionless length of $550D$. Taking the {entrance length}  into account the remaining measurement length is {$350D$}. Here the measurement length is defined as the distance between the perturbation and the measurement point and the puff formation time is observed to be $t_0 \approx 100 D/U_s$ in agreement with \cite{Hof08}. In the second case the pipe was composed of acrylic tubes (inner diameter $D=7.18 \pm 0.02 ~\textrm{mm}$) and had a total length of $12~\textrm{m}$, leading to a measurement length of $1300D$. In the third case the pipe was composed of glass tubes (inner diameter $D=4 \pm 0.01 ~\textrm{mm}$) and had a total length of $10~\textrm{m}$ and a measurement length of $2250D$. In these cases pipe segments were joined by Perspex connectors which guarantee a seamless fit. The pipe segments were positioned and carefully aligned on a long aluminum profile. The pipe is connected via a trumpet shaped convergence section to a reservoir (see FIG. \ref{fig:facility}). The rear end of the pipe is connected to a $1.2~\textrm{m}$ long piston (Pneumax) with a diameter of $40~\textrm{mm}$ that pulls the water through the pipe. The piston sits on a separate support and is coupled to the pipe via a short piece of semi-flexible tubing to dampen vibrations from the mechanical drive. The piston plunger is connected to a linear traverse (HepcoMotion PSD120) moved by a stepper motor (Dunkermotoren BG65PI). When run in steady motion the flow remains laminar up to $Re=9000$.

The piston speed was accurately controlled via a PC and was sinusoidally modulated giving rise to a pulsating flow ($u(t) = U_s + U_o\cdot \textrm{sin}(2\pi f \cdot t)$). For the entire parameter regime under investigation the pipe flow was laminar unless being disturbed. 
To probe the quality of the pipe facility particle image velocimetry measurements were carried out for a pulsating flow at $Re_s=2000$, $\alpha=5$ and $A=0.4$ in the absence of any perturbation. The velocity profiles measured during various phases of the cycle are in excellent agreement with the analytical solution for laminar pulsating pipe flow. Disturbances could be added to the flow by a brief injection (using an electronically regulated valve) of fluid through a $1~\textrm{mm}$ hole in the pipe wall. The ratio between the injection and the pipe flow in volume flow rate is about $2\%$ which is sufficiently strong to generate a single puff at the injection point \citep{Hof03}. This perturbation was located $150D$ downstream from the inlet where the laminar flow can be regarded as fully developed \citep{Florio68, Gerrard71}.

\begin{table}
	\begin{center}
		\resizebox{0.5\columnwidth}{!}{
			\begin{tabular}{cccccc}
				$L (D)$ & $A$ & $\alpha$ & $T_{\textrm{inj}}$ ($\% T$) & $\Phi_{\textrm{inj}}$ ($\times 2\pi$) \\
				\hline
				350  & 0.4 & 3.96  & 6.25 & 0.25 $\sim$ 0.3125 \\
				350  & 0.4 & 5.59  & 25   & 0.25 $\sim$ 0.50 \\
				350  & 0.4 & 7.22  & 50   & 0.25 $\sim$ 0.75 \\
				350  & 0.4 & 8.84  & 50   & 0.25 $\sim$ 0.75 \\
				350  & 0.4 & 10.21 & 50   & 0.25 $\sim$ 0.75 \\
				350  & 0.4 & 11.41 & 50   & 0.25 $\sim$ 0.75 \\
				350  & 0.4 & 12.51 & 100  & 0 $\sim$ 1 \\
				350  & 0.4 & 14.44 & 100  & 0 $\sim$ 1 \\
				350  & 0.4 & 16.14 & 100  & 0 $\sim$ 1 \\
				350  & 0.4 & 17.69 & 100  & 0 $\sim$ 1 \\
				350  & 0.4 & 21.66 & 100  & 0 $\sim$ 1 \\
				350  & 0.2 & 7.22  & 50   & 0.25 $\sim$ 0.75 \\
				350  & 0.2 & 8.84  & 50   & 0.25 $\sim$ 0.75 \\
				350  & 0.6 & 8.84  & 50   & 0.25 $\sim$ 0.75 \\
				1300 & 0.4 & 2.48 & 2.5   & 0.25 $\sim$ 0.275 \\
				1300 & 0.4 & 3.62 & 2.5   & 0.25 $\sim$ 0.275 \\
				1300 & 0.4 & 3.92 & 2.5   & 0.25 $\sim$ 0.275 \\
				1300 & 0.4 & 6.20 & 15    & 0.25 $\sim$ 0.40 \\
				1300 & 0.2 & 5.55 & 15    & 0.25 $\sim$ 0.40 \\
				1300 & 0.4 & 5.55 & 10    & 0.25 $\sim$ 0.35 \\
				1300 & 0.55 & 5.55 & 10   & 0.25 $\sim$ 0.35 \\
				1300 & 0.2 & 3.92 & 7.5   & 0.25 $\sim$ 0.325 \\
				1300 & 0.4 & 3.92 & 7.5   & 0.25 $\sim$ 0.325 \\
				1300 & 0.6 & 3.92 & 5     & 0.25 $\sim$ 0.30 \\
				1300 & 0.7 & 3.92 & 3     & 0.25 $\sim$ 0.28 \\
				2250 & 0.4 & 1.50 & 1.0  & 0.25 $\sim$ 0.26 \\
				2250 & 0.4 & 2.00 & 1.5  & 0.25 $\sim$ 0.265 \\
				2250 & 0.4 & 2.50 & 2.5  & 0.25 $\sim$ 0.275 \\
				\hline
			\end{tabular}
		}
	\end{center}
	\caption{The parameters of pulsatile pipe flow experiments. $T_{\textrm{inj}}$ is the perturbation duration normalized by the period, and $\Phi_{\textrm{inj}}$ is the perturbation phase covered in sinusoidal mass flow rate.} \label{tab:parameter}
\end{table}

To ensure that the perturbation resulted in a single puff (the length of a puff is about $20D$ \citep{Samanta11}), the duration of injection is adjusted for each $\alpha$ to cover a specific phase of the sinusoidal motion, as demonstrated in FIG. \ref{fig:facility}. Variations in the phase of the injection turned out to have no noticeable effect on the transition threshold. The injection duration on the other hand turned out to be relevant because long durations resulted in multiple puffs. To avoid this and to ensure that only single puffs were triggered injection times were limited to a maximum of $20 D/U_s$. For visualization purposes, the water was seeded with particles (fishsilver), and a light sheet and a camera (MatrixVision BlueFox 121G) were positioned a certain distance $L$ downstream from the perturbation point, as shown in FIG. \ref{fig:facility}. The resulting images allow for a straightforward distinction between turbulent and laminar flows (see FIG. \ref{fig:facility} (e, f)). Puffs could be readily detected by a change in the average grey scale level at the pipe centre or by monitoring spatial fluctuations (computing the root mean square). Before each run the water temperature {was} monitored with a calibrated Pt100 probe to determine possible changes in the fluids viscosity. The flow rate was then automatically adjusted to ensure the desired value of $Re_s$.
While our initial studies were carried out in a pipe  with an effective measurement length (i.e. total observation time) of $350D$, at lower Womersley numbers the advection over one period exceeded this length and those measurements were carried out in the longer pipes ($1300D$ and $2250D$ respectively). It should be noted that the survival probabilities obtained in the three pipes weakly depend on the pipe length. In the longer pipes puffs will only survive over the longer measurement length at somewhat higher $Re$ where their lifetimes are larger. The threshold estimates obtained in the longer pipes are therefore closer to the actual critical point where turbulence eventually becomes sustained. However, these differences are small ($<10 \%$ for steady pipe flow) compared to the change of the transition threshold with Womersley number ($>50 \%$). The experimental parameters are summarized in table \ref{tab:parameter}.

\section{Results}

\begin{figure}
\centering
\includegraphics[width = 0.75\textwidth, trim=0cm 0.25cm 0.25cm 0cm]{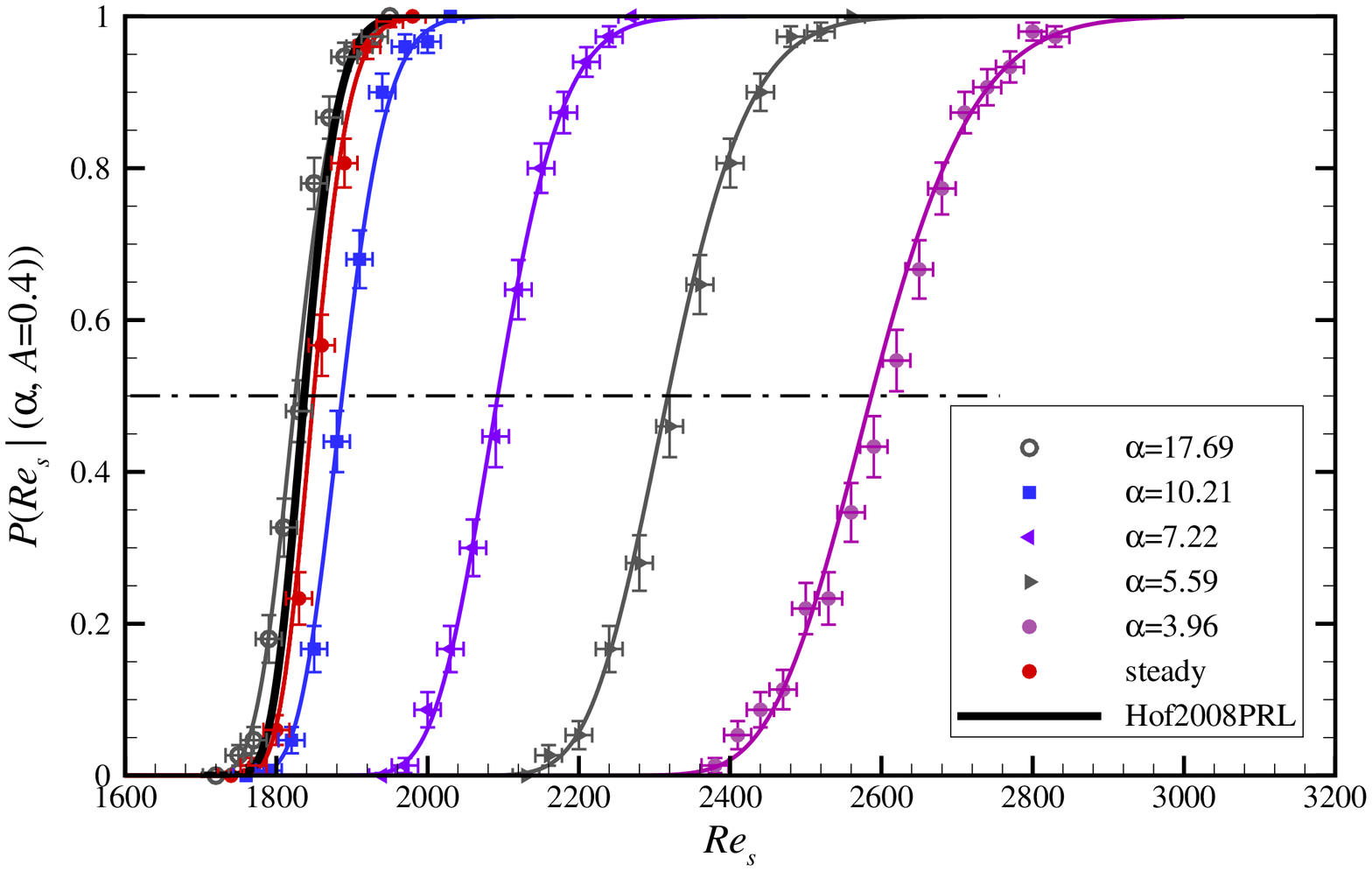}\\
\includegraphics[width = 0.75\textwidth, trim=0.5cm 0cm 0cm 0.5cm]{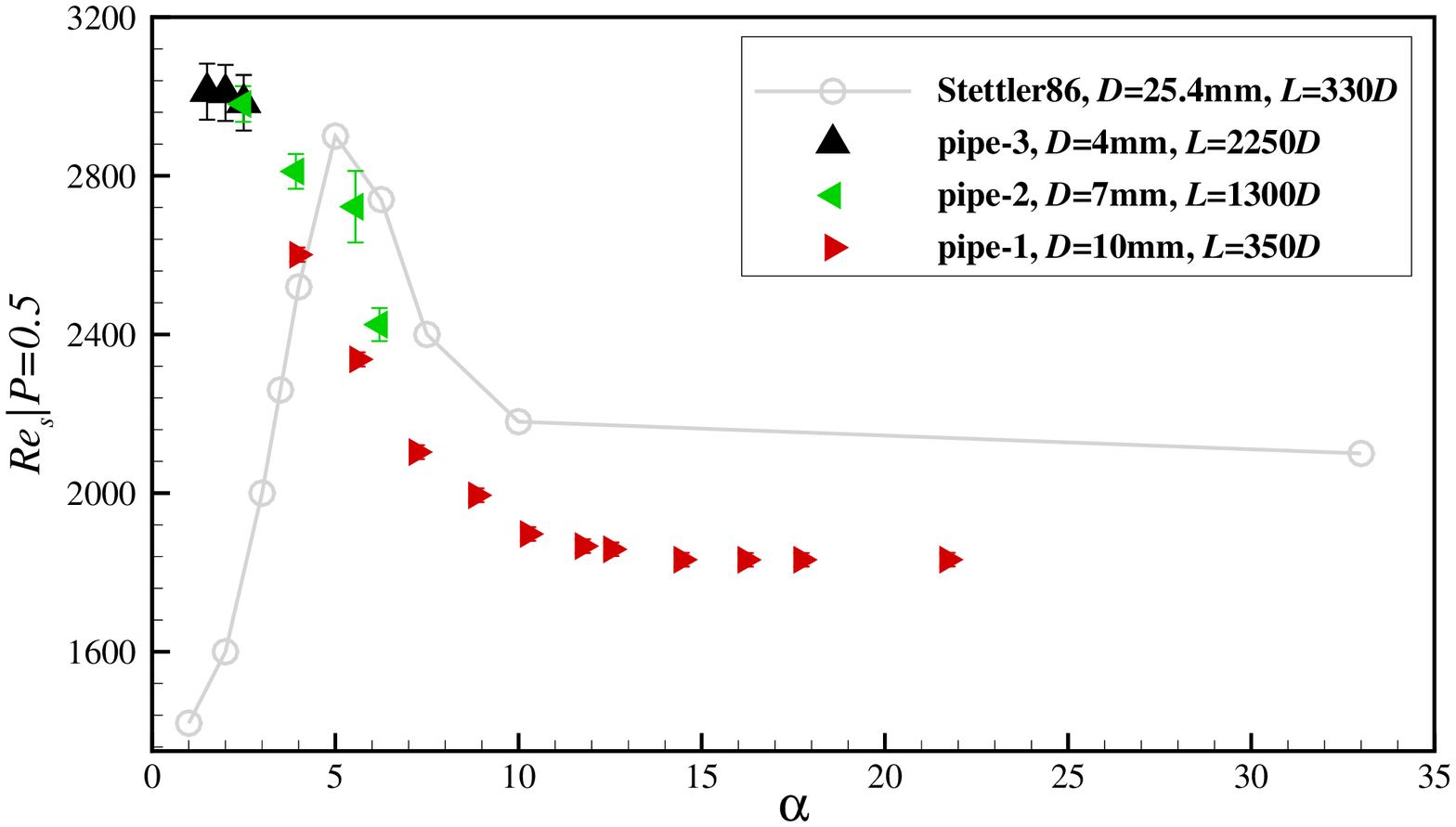}\\
\caption{\label{fig:frequency_curve} (Color online) (Top) The survival probability of {individual puffs, measured $350 D$ downstream from the perturbation point. Plotted are the survival probabilities as a function of $Re_s$ for various pulsation frequencies and a fixed amplitude of $A=0.4$.} The corresponding lines indicate super-exponential fits to the respective data sets. The black curve shows the lifetime scaling for steady pipe flow from \citet{Hof08}. For clarity not all measured data sets are plotted. The dash-dot line marks $P=0.5$. (Bottom) The Reynolds number for which $P=0.5$ is plotted as a function of Womersley number. Measurements are carried out for three different lengths, corresponding to three different observation times $t=350$, $1300$, and $2250$ (for all cases puff advection speeds are close to 1). Subtracting the puff formation time $t_0 \approx 100D/U_s$ the obtained Reynolds number values then map out the thresholds where puff lifetimes reach 250, 1200 and 2150 advective time units respectively. The results of \citet{Stettler86} are shown in grey for comparison. The error bars for $P$ are $95\%$ confidence intervals of a binomial distribution as a function of $(N_s,N_t)$ obtained with the Wilson method \citep{Brown01}, whereas the other error bars show the uncertainties of the corresponding quantities from measurements.}
\end{figure}

To study the effect of pulsation on transition, experiments were carried out for Womersley numbers $1.5 \leq \alpha \leq 22$, amplitudes $0 \leq A \leq 0.7$ and Reynolds numbers $Re_s<3500$. In this parameter regime flows were found to be laminar (FIG. \ref{fig:snapshot} left column) unless being perturbed and hence the laminar flow appears to be linearly stable. For large enough $Re_s$ however a sufficiently strong perturbation resulted in a localised patch of turbulence (shown in FIG. \ref{fig:snapshot} right column). To determine the transition threshold and its dependence on flow pulsation a detailed parameter study was carried out.
For fixed combinations of $(\alpha, A)$, $Re_s$ was varied to find the regime where sufficiently strong perturbations first result in turbulent puffs. Next we determined the probability for puffs to survive over a fixed distance (the measurement distance is {$350D$} for pipe 1, {$1300D$} for pipe 2 and {$2250D$} for pipe 3) as a function of $Re_s$. For each $Re_s$ $150$ runs were carried out to obtain statistically converged survival probabilities. For individual puffs the survival probability is defined as $P(Re_s| (\alpha, A))=N_s/N_t$, where $N_s$ is the number of survival cases and $N_t$ the total number of runs.

In steady pipe flow, the probability for a puff to survive for a time $t$ is only a function of Reynolds number, $P(Re,t) = \textrm{exp}[-(t-t_0)/\tau(Re)]$,
where $t$ is the time since the start of the experiment, $\tau$ is the characteristic lifetime of a puff and $t_0$ represents the time for the initial formation of a puff (which we measured to be $t_0 \approx 100D/U$).
For a steady flowrate (red circles in FIG. \ref{fig:frequency_curve}, top) survival probabilities measured in the present piston driven set up (pipe 1) are in excellent agreement with previously reported values (black solid curve in FIG. \ref{fig:frequency_curve}) for pipes operated at a constant pressure head \citep{Hof08}. The survival probability increases with $Re$ following a S-shaped curve (see e.g. \citet{Hof06, Hof08}) and this scaling is indicative for the transient nature of individual puffs: A survival probability of $P=1$ is only approached asymptotically.
{Next we investigated the effect of pulsation frequency for a fixed pulsation amplitude of $A=0.4$, and the measured probabilities are shown in FIG. \ref{fig:frequency_curve}.  Survival probabilities in pulsating flow equally follow S-shaped curves owing to the transient nature of individual puffs. For small Womersley numbers S-curves are considerably shifted to the right and hence much higher $Re_s$ are required to observe puffs of appreciable lifetimes. For large Womersley numbers on the other hand puff survival probabilities approach the steady pipe flow case and the transition point appears to be unaffected by flow pulsation. To better quantify this trend we determine (for each $\alpha$) the Reynolds number where $P(Re_s, t-t_{0})=0.5$, i.e., the Reynolds number where $50\%$ of the puffs survive up to the measurement point. As shown in FIG. \ref{fig:frequency_curve} (bottom panel) for $\alpha > 12$ $50\%$ survival probabilities are reached at the same Reynolds number as in steady flow {($Re \approx 1860$ for $L=350D$)}. For $\alpha < 12$ however transition is delayed and significantly higher Reynolds numbers ({$Re_s=2960$} for $\alpha=2.5$) are necessary to obtain puffs of the same lifetimes as in the steady case.} Measurements of puff lifetimes at lower Womersley numbers had to be carried out in the longer pipe set ups, in order to ensure that puffs experience a full oscillation cycle before they exit the pipe. The data sets (FIG. \ref{fig:frequency_curve}, bottom) taken in the two longer pipes show the same general trend, transition thresholds keep increasing with decreasing Womersley number. In contrast to earlier studies of this parameter regime that reported transition to occur earlier than in steady flow, the transition threshold continues to increase and there is no sign of a reversal of this trend. The rate of increase however slows down  for $ \alpha < 2.5$.
\begin{figure}
\centering
\includegraphics[width = 0.75\textwidth, trim=0.5cm 0cm 1cm 0.5cm, clip]{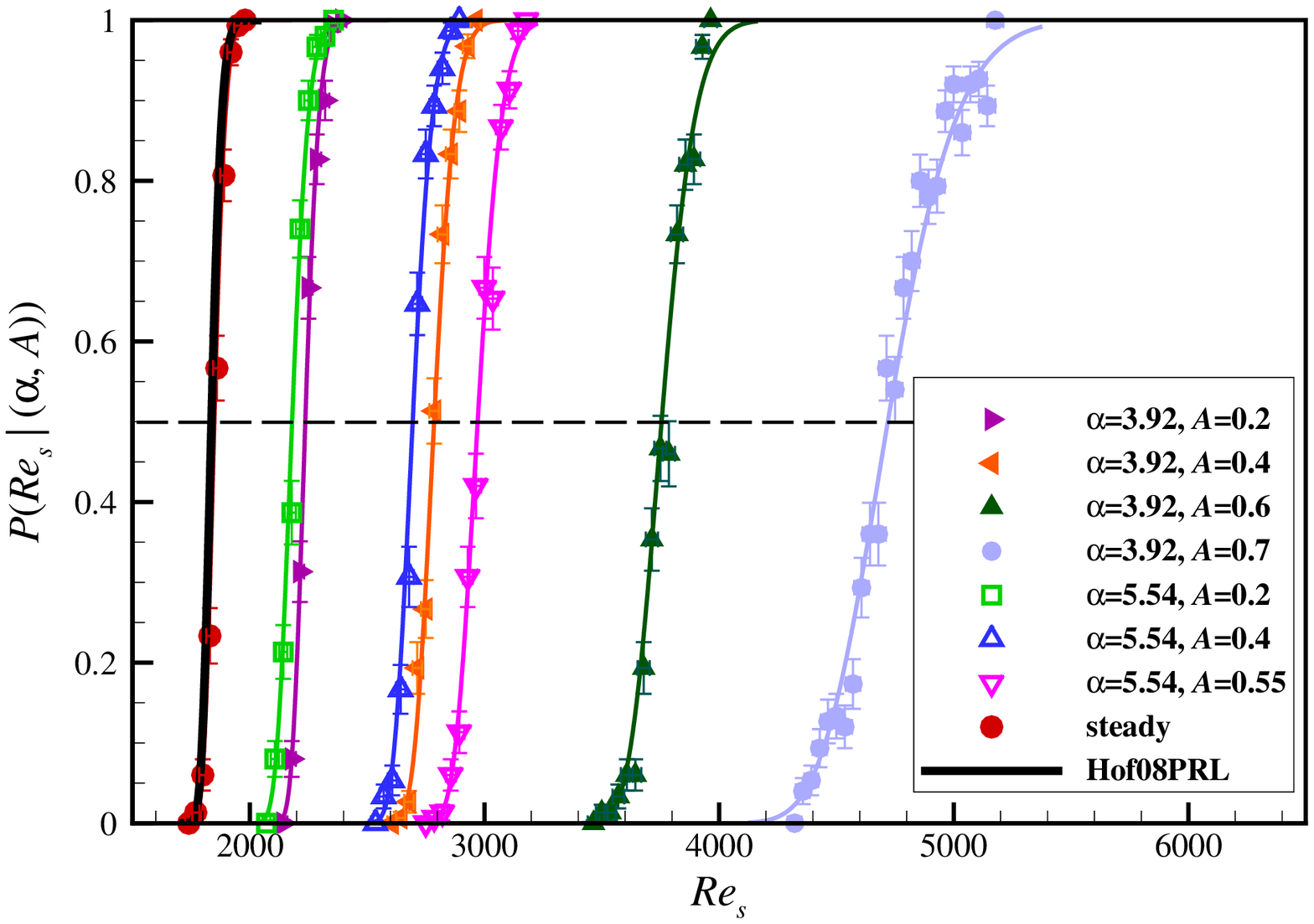}\\
\includegraphics[width = 0.75\textwidth,trim=1cm 0cm 0.25cm 1cm,clip]{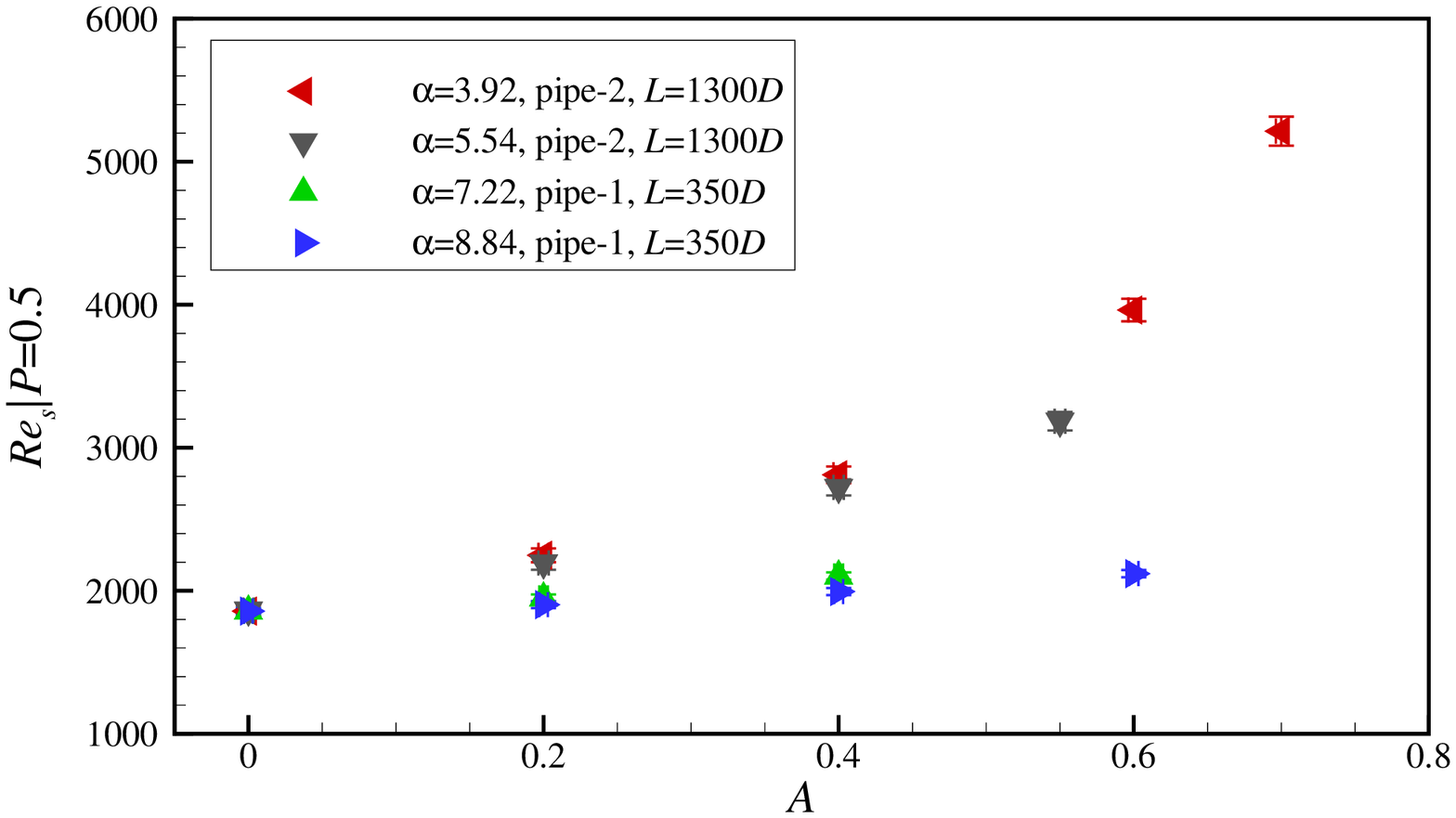}
\caption{\label{fig:amplitude} (Color online) {(Top) The survival probability of individual puffs measured for varying pulsation amplitudes. The dashed line marks $P=0.5$. (Bottom) The Reynolds number threshold values ($50\%$ survival probabilities) are plotted as a function of the pulsation amplitude. The error bars for $P$ are $95\%$ confidence intervals of a binomial distribution as a function of $(N_s,N_t)$ obtained with the Wilson method \citep{Brown01}, whereas the other error bars show the uncertainties of the corresponding quantities from measurements.}}
\end{figure}

In addition to the role of the Womersley number we also tested the influence of the pulsation amplitude on transition. Here the pulsation frequency was fixed and $A$ was varied. Again we measured puff survival probabilities and from the S-shaped curves (shown in FIG. \ref{fig:amplitude}) $Re_s|{P=0.5}$ was determined. As shown in FIG. \ref{fig:amplitude} (top panel)  the characteristic S-shapes are shifted considerably to higher $Re_s$ and the transition delay increases with pulsation amplitude. We observed that for all four frequencies investigated $Re_s|{P=0.5}$ increases monotonically with pulsation amplitude (FIG. \ref{fig:amplitude}, bottom panel) and this transition delay is most prominent for low frequencies while at large frequencies the increase with amplitude is only moderate.

\section{Discussion}
The often conflicting results reported in previous studies of the onset of turbulence in pulsatile flows reflect the difficulty in quantifying the transition process in linearly stable flows, in particular if there are several control parameters. Puffs are the first turbulent structures to arise in the subcritical regime of pulsatile pipe flow, and lifetimes of turbulent puffs provide a natural and very accurate measure of the transition threshold, because small changes in the governing parameters can change mean lifetimes by several orders of magnitude. Our detailed studies of puff survival probabilities show that the transition threshold overall decreases with $\alpha$ despite the fact that most earlier studies \citep{Nerem72a,Nerem72b,Falsetti83,Peacock98} reported the opposite trend. A main difference is that while in the present study turbulence is triggered actively for each measurement, in their case the flow was left to its own devices without any knowledge or control of the perturbation levels. Turbulence in their studies may have hence existed at considerably lower parameters but simply may have not been triggered, lacking sufficiently strong disturbances. If perturbation levels to trigger turbulence depend on $\alpha$ or if the noise levels present in the experiments change with $\alpha$, experiments lacking an active triggering mechanism become very difficult to interpret.

Our observations that the transition threshold remains unchanged in the large frequency limit ($\alpha>12$) agrees very well with the earlier studies by \citet{Trip12, Stettler86}. Note that in these two studies turbulence was actively triggered (just like in our study). Once the Womersley number decreases below $10$ however, puff lifetimes decrease notably and transition is delayed to larger $Re_s$. While this regime was not covered by \citet{Trip12}, \citet{Stettler86} had also reported transition delay for $5<\alpha <10$ although threshold values between their and our study differ by $\sim 20\%$. In contrast to our study they found however that for $\alpha <5$ the transition threshold decreases. On the other hand it should be noted that the pipe length used in their study is insufficient to investigate such low Womersley numbers: turbulent structures created at the inlet can only experience part of the oscillation cycle as they were flushed out of the pipe before a full cycle finished. The average $Re$ that turbulence experiences over such a partial cycle then depends on the phase at which turbulence is created, which will necessarily cause deviations from the true transition threshold. In particular turbulent structures that {traverse} the pipe during the faster half of the cycle ($Re(t)>Re_s$) experience a larger effective Reynolds number and will reach the end of the pipe, although in a longer pipe they would have decayed during the slower half of the cycle ($Re(t)<Re_s$).

When varying pulsation amplitudes, we observed that the transition delay increases monotonically. This delay with increasing amplitude is observed for all frequencies studied here, but it is much more pronounced at low $\alpha$.  Earlier studies had reported a transition delay in this regime \citep{Stettler86, Sarpkaya66}, but they reported that at larger amplitudes the transition moved back to lower Reynolds number, e.g., \citet{Sarpkaya66} reported that for fixed $\alpha(=5.8)$ as the pulsation amplitude increases from $A=0.5$ to $A=0.6$ $Re_s=3800$ decreases to $Re_s=3000$, which is not confirmed in our study. It should however be noted that for larger amplitudes other instabilities may arise and the present study only considers the subcritical transition to turbulence via puffs.

\begin{figure}
\centering
\includegraphics[width = 0.8\textwidth]{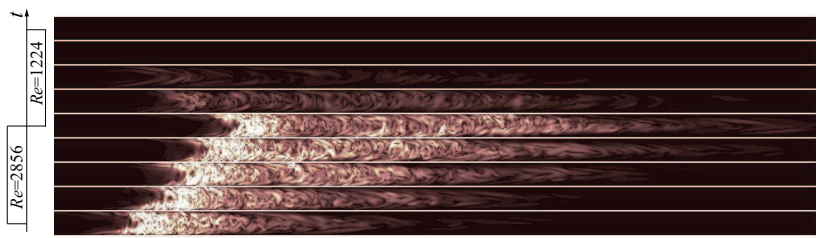}\\
\caption{\label{fig:DNS} (Color online) {Direct Numerical Simulation of a puff in a square-wave motion at $Re_s=2040$ with $A=0.4$. {From bottom to top, each snapshot shows the contour of the streamwise vorticity normalised by mean velocity of $Re_s$, from $t=0$ to $t=T$ with an equal time interval of $T/8$.}}}
\end{figure}

The surprising observation that in the small Womersley number limit the upward shift of the transition point persists can be understood when looking at the growth and decay processes of turbulence. This argument is most easily illustrated for the simpler case of a square wave pulsation, i.e. where the Reynolds number jumps between two levels. We have carried out a direct numerical simulation setting the amplitude to $40\%$ and the mean Reynolds number to $2040$ (corresponding to the critical point of steady pipe flow, \cite{Avila11}).
The simulation was carried out in a $100D$ long pipe with periodic boundary condition imposed in the axial direction using the hybrid finite difference-Fourier spectral method described in \citet{Willis2009}. The spatial resolution is $64\times 96\times 2560$ grid points in radial, azimuthal and axial directions. At $Re=2856$, i.e. the faster half, we perturbed the flow with a puff that we simulated at $Re=2040$ in the same pipe. Then the puff starts to grow turning into a slug. At the jump to the lower cycle at $Re=1224$, we abruptly changed the mass flux (mean flow speed)  while keep the turbulent fluctuations unchanged. This mimics the situation in experiment in which the mass flux is changed quickly by, e.g., abruptly adjusting the speed of the driving piston. An oscillation period of $120$ advective time units (corresponding to a Womersley number of $5$) was simulated.

As shown in FIG. \ref{fig:DNS} during the faster half of the square-wave cycle ($Re=2856$, bottom half of FIG. \ref{fig:DNS}) turbulence rapidly expands. After half of the oscillation period ($60$ time units) $Re$ changes to $1224$, and turbulence abruptly decays (top half of FIG. \ref{fig:DNS}). A key difference is that in the faster part of the cycle the turbulent patch expands outwards from its interfaces, whereas at low $Re$ turbulence collapses throughout (rather than receding from its interfaces). The decay is much faster and clearly dominates over the growth process when viewed over one full pulsation period.  This imbalance between the decay and growth implies that in the low Womersley number regime turbulence can only become sustained at significantly larger $Re$ than in steady pipe flow, in qualitative agreement with the experimental observations.

\begin{figure}
\centering
\includegraphics[width = 0.6\textwidth, trim=0cm 0cm 0cm 0cm, clip]{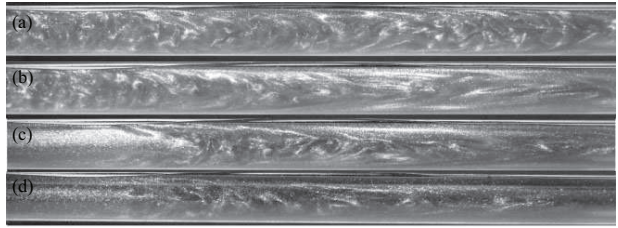}\\
\caption{\label{fig:quasi_steady} {Flow structures recorded at an instantaneous Reynolds number $Re(t)=1800$ for various $\alpha$: (a) $\alpha=8$, (b) $\alpha=5$, (c) $\alpha=3$; (d) steady pipe flow at $Re=1800$. For the pulsating flows (a)-(c) all other parameters are identical $Re_s=3000$ and $A=0.4$ and the images were recorded at the minimum of the cycle (i.e. at $Re=1800$). At $\alpha=8$ turbulence forms an elongated slug, despite the low instantaneous Reynolds number ($1800$). For $\alpha=3$ on the other hand the observed flow structure is a puff in close correspondence to puffs in steady flow of the same $Re$ (d).}}
\end{figure}

For sinusoidal pulsation the collapse of turbulence will be less abrupt than in above example due to the smoother Reynolds number decrease, nevertheless it will equally dominate the prior growth if Womersley number are small. In particular we would expect that here, for sufficiently slow Reynolds number variations,  expanding turbulent structures (i.e. slugs) collapse to puffs when $Re$ decreases to the puff regime. To investigate if at sufficiently low Womersley numbers this proposed structural adjustment of turbulence indeed occurs we compare flows at $\alpha=$ 8, 5 and 3. In all three runs the mean Reynolds number is equal to $3000$ while the pulsation amplitude is $0.4$. In each case the laminar flow is perturbed at $t=T/4$ where the flow speed is at its maximum ($Re= 4200$). We then follow the evolving turbulent flow downstream and visualise the flow field at $t=3T/4$, i.e. at the minimum flow speed ($Re=1800$). The main difference among the three runs is the rate at which the flow speed sinusoidally adjusts between the maximum and minimum Reynolds number: for $\alpha=8$ the time that elapses corresponds to $(T/2=)37.5 D/U_s$, while for $\alpha=3$ the Reynolds number adjustment is spread out over a much longer time of $250 D/U_s$.
At $\alpha=8$ turbulence takes the form of a slug-like structure (FIG. \ref{fig:quasi_steady} (a)), whereas in a steady flow at the same Reynolds number ($1800$) slugs are unstable and only much shorter puffs can be found (see FIG.\ref{fig:quasi_steady} (d)). The fact that in this case  (FIG. \ref{fig:quasi_steady} (a)) an elongated turbulent structure is found despite the low instantaneous Reynolds number must be attributed to the relatively fast drop from a much higher Reynolds number. The Reynolds number change appears to be too fast for turbulence to adapt to the structure characteristic for the instantaneous speed. While puffs are prone to decay, slugs are stable and hence turbulence at $Wo=8$, $Re_s=3000$ is sustained and well above the transition threshold. At the intermediate Womersley number (FIG. \ref{fig:quasi_steady} (b)) the slug observed at the minimum of the cycle (i.e. $Re=1800$) is shorter than at $\alpha=8$ but still larger than the structures would be under steady flow conditions at this instantaneous $Re$. Finally for $\alpha=3$ at the cycle minimum turbulence has reduced to a puff (FIG. \ref{fig:quasi_steady} (c)), structurally equivalent to puffs in steady pipe flow at this Reynolds number (FIG. \ref{fig:quasi_steady} (d)). Although during the faster part of the cycle turbulence had spread considerably (not shown) once $Re$ drops {sufficiently (here shown for $Re=1800$)}, the extended slug (in this example $> 100 D$ at its maximum length) collapses and in this case only a single puff (at the position of the slug's trailing edge) remained. It hence appears that at low Womersley numbers where Reynolds number changes are much slower, turbulent structures can more readily adapt to the instantaneous Reynolds numbers. As Womersley numbers further decrease one may expect that eventually turbulent structures follow the very slow changes of Reynolds number in a quasi steady fashion. In this case the survival of turbulence would be decided over the low Reynolds number phase where turbulence consists of distinct puffs with finite lifetimes. In this limit it should then be possible to estimate the transition threshold based on puff lifetimes by integrating over the corresponding steady flow values for the corresponding Reynolds numbers. For instance, the survival probability $P$ is obtained by integration over the instantaneous decay rates which are $\kappa(t) \delta t$ at time $t$, where $\kappa$ is the escape rate \citep{Avila10}:

\begin{equation}
P = \textrm{exp}\left[-\int_{t_1}^{t_2}	  \kappa( Re(t)) \delta t \right].
\end{equation}

\begin{figure}
\centering
\includegraphics[width = 0.8\textwidth, trim=0cm 0cm 0cm 0cm, clip]{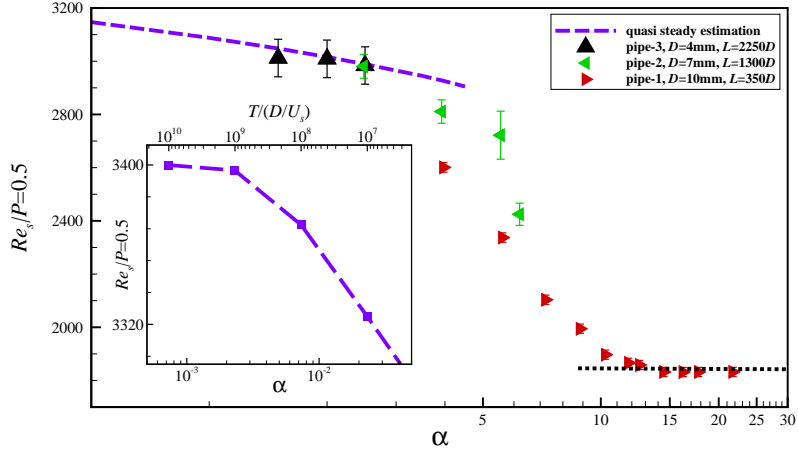}\\
\caption{\label{fig:estimation} (Color online) Estimates of the transition threshold, obtained by determining the Reynolds number value where puffs have a $50\%$ survival probability (i.e. $Re_s(\alpha) | P=0.5$) over the respective pipe length $L$ of the three different pipe set ups. For comparison the (purple) dashed curve shows the $50\%$ survival probability over a single period obtained from the quasi steady assumption. {For $\alpha \rightarrow 0$ the curve asymptotes towards an upper bound ($Re_s=3400$ for $A=0.4$) (see inset).} The black dotted line marks the $50\%$ survival probability for $L=350D$ (same as red circles) for steady pipe flow.}
\end{figure}

Here $[t_1,t_2]$ is the time interval of the pulsation cycle for which $Re \leq 2040$ and the escape rate $\kappa$ is equal to the inverse of the median lifetime $\tau(Re)$ which for steady pipe flow is $\tau=\textrm{exp}(\textrm{exp}(0.00556 Re -8.5))$ \citep{Avila11}.
For fixed values of $\alpha$ we then determine from equation 4.1 the mean Reynolds number ($Re_s$) at which $50 \% $ of the puffs survive over one period, i.e. $Re_s|P=0.5$. When decreasing the Womersley number while keeping $Re_s$ fixed, the length of the time interval $[t_1,t_2]$ increases and consequently the survival probability (equation 4.1) decreases. In order to retain the same survival probability (i.e. $P=0.5$) at lower $\alpha$ the mean Reynolds number has to go up. Therefore the quasi steady assumption predicts monotonically increasing threshold values $Re_s|P=0.5$ for decreasing $\alpha$. This at least qualitatively agrees with the increasing trend observed in experiments. To test the quantitative agreement the results of the quasi steady analysis are shown for comparison in FIG. \ref{fig:estimation} (dashed curve). For $\alpha < 2.5$ they indeed closely capture the values measured in the pulsating flow experiment. It would appear that for the long period times (and hence the slow Reynolds number variations at these low Womersley numbers) flows indeed can be considered as quasi steady.
The threshold increase predicted by the quasi steady assumption is bounded from above because decay rates become zero at $Re=2040$ (as suggested by \cite{Avila11}). If $Re$ remains above $2040$ for the entire cycle turbulence ceases to decay. Determining the $P=0.5$ threshold from equation (4.1) for very low Womersley number values $Re_s|P=0.5$ indeed levels off at a value of $3400$, as shown in the inset of FIG. \ref{fig:estimation}. Here the minimum Reynolds number of the cycle is $3400(1-A)=2040$, the point where decay rates become zero. Unfortunately due to the very long periods at these low Womersley numbers an experimental verification of this predicted asymptotic value is not possible.

\section{Conclusion}

The transition to turbulence in pulsating pipe flow can be separated into three regimes: 1. In the high frequency limit $\alpha \gtrsim 12$ the transition threshold is unaffected by flow pulsation. Presumably here flowrate changes are too fast for turbulence to react and hence turbulence becomes sustained when {\it{the average Reynolds number}} is equal to the steady state threshold. 2. For low Womersley numbers ($\alpha \lesssim 2.5$) the Reynolds number changes are sufficiently slow for turbulent structures to adjust to the instantaneous Reynolds number and turbulent lifetimes can be predicted by a quasi steady assumption. Much of the expansion of turbulence occurring during the faster part of the cycle is erased once $Re$ drops and turbulence reduces to puffs ($Re \lesssim 2300$ \cite{Barkley15}). This low Reynolds number interval is critical for the survival of turbulence and the transition threshold is determined by the puff decay rates in this interval. Therefore transition thresholds are considerably larger than in high Womersley number flows and equally transition is delayed compared to steady flow. As suggested by the quasi steady analysis, in the low Womersley number limit {\it{the minimum Reynolds number}} has to be above threshold for turbulence to be sustained. 3. In the intermediate Womersley regime, i.e. $2.5 \lesssim \alpha \lesssim 12$ the transition threshold adjusts smoothly between the two limits.

\begin{acknowledgments}
{D.X. gratefully acknowledges the support from Max Planck Society and Humboldt Foundation (3.5-CHN/1154663STP) and IST Austria. B.H. acknowledges funding from the European Research Council under the European Unions Seventh Framework Programme (FP/2007-2013)/ERC Grant Agreement 306589.}
\end{acknowledgments}

\bibliographystyle{jfm}
\bibliography{jfm16_arxiv}

\end{document}